%% file: main.tex
\documentclass[10pt]{cai}

\begin{document}
\def\conferenceyear{2026}
\volumeheader{37}{0}
\begin{center}

\title{Generative Latent Alignment for Interpretable Radar Based Occupancy Detection in Ambient Assisted Living}
\maketitle

\thispagestyle{empty}

\begin{tabular}{cc}
Huy Trinh\upstairs{\affilone,\affiltwo,*}
\\[0.25ex]
{\small \upstairs{\affilone} University of Waterloo} \\
{\small \upstairs{\affiltwo} Vector Institute} \\
\end{tabular}
  
\emails{
  \upstairs{*}h3trinh@uwaterloo.ca 
}
\vspace*{0.2in}
\end{center}

\begin{abstract}
In this work, we study how to make mmWave radar presence detection more interpretable for Ambient Assisted Living (AAL) settings, where camera-based sensing raises privacy concerns. Our Generative Latent Alignment (GLA) framework combines a lightweight convolutional variational autoencoder with a frozen CLIP text encoder to learn a low-dimensional latent representation of radar Range--Angle (RA) heatmaps. The latent space is softly aligned with two semantic anchors corresponding to ``empty room'' and ``person present'', and Grad-CAM is applied in this aligned latent space to visualize which spatial regions support each presence decision. On our mmWave radar dataset, we qualitatively observe that the ``person present'' class produces compact Grad-CAM blobs that coincide with strong RA returns, whereas ``empty room'' samples yield diffuse or absent evidence. In addition, we conduct an ablation study with unrelated text prompts, which degrades both reconstruction and localization, suggesting that radar-specific anchors are important for meaningful explanations in this setting.
\end{abstract}

\begin{keywords}{Keywords:}
Ambient Assisted Living, explainable AI, Grad-CAM, mmWave radar, semantic alignment, variational autoencoder.
\end{keywords}
\copyrightnotice

\section{Introduction}
\input{sec/introduction} 

\section{Related Work}
\input{sec/related_work}

\section{Methodology}
\label{sec:methodology}
\input{sec/method}

\section{Experiments and Results}
\input{sec/experiments_and_result}
\printbibliography[heading=subbibintoc]

\end{document}

%% file: sec/introduction.tex
With the proliferation of next-generation  wearable and smart home devices, embedded intelligence is pushed into the physical environment where we work and live. Ambient Assisted Living (AAL)  aims to support older adults' autonomy and safety through continuous monitoring of human presence and activities~\cite{AAL, s21103549}. 
While cameras are effective for activity recognition, they suffer from major privacy issues.
Radio-frequency (RF) sensing, in particular millimeter-wave (mmWave) radar, offers an appealing alternative due to its robustness to lighting conditions, privacy preservation and tolerance to occlusion~\cite{10554983}.
AAL systems based on deep learning already play an important role in presence detection, fall detection, and occupancy monitoring. However, this rapid progress has also highlighted a significant challenge: the lack of interpretability~\cite{Nahar}.
In safety-critical healthcare contexts, such limited transparency makes it difficult for clinicians and operators to trust or validate the behaviour of deployed models.
To address this gap, we propose a Generative Latent Alignment (GLA) framework designed to bridge the representation gap between abstract radar signals and human-understandable concepts.
Our method couples a lightweight convolutional variational autoencoder (VAE) on radar Range--Angle (RA) maps with a simple text-anchored latent alignment head and latent-space Grad-CAM. We validate GLA on a binary radar presence dataset, showing that it produces class-conditional Grad-CAM visualizations that highlight the RA regions used for the decision and are qualitatively consistent with the underlying radar structure. 

%% file: sec/related_work.tex
Recent literature in radar signal processing has shifted from purely discriminative models to generative approaches to handle data scarcity and noise~\cite{11171388}. Variational Autoencoders (VAEs) have been successfully applied to FMCW radar data to learn low-dimensional representations of Range-Doppler and Range-Azimuth feature maps, showing that reconstruction objectives help the model learn robust features that generalize better than supervised learning alone~\cite{10149738}.
However, while these models learn structural texture of the environment, their latent variables are cryptic and not necessarily correspond to human-understandable concepts.
To bridge the gap between sensor data and human understanding, recent work aligns radar or other sensor embeddings with large vision–language models such as CLIP~\cite{radford2021learningtransferablevisualmodels} and LLM back-end, enabling zero-shot activity understanding and natural-language querying of mmWave radar scenes~\cite{lai2025radarllmempoweringlargelanguage}.
Otto et al.~\cite{s25144467} propose generating descriptive text from radar classifiers using adversarial training. 
In parallel, explainable-AI efforts for sensing propose saliency-based methods, rule-based models to interpret sensor signals and room-level presence decisions, but these typically operate post-hoc on discriminative models rather than shaping the generative latent space itself~\cite{11071277,rafique2025interpretable}. Our work builds on these directions by coupling a lightweight radar VAE with simple CLIP-based text anchors and latent Grad-CAM, aiming for class-conditional evidence maps that are both grounded in RA structure and semantically tied to radar domain knowledge without requiring large language models or complex decoder heads.

%% file: sec/method.tex
\begin{figure}[ht]
    \centering
    \includegraphics[scale=0.32,]{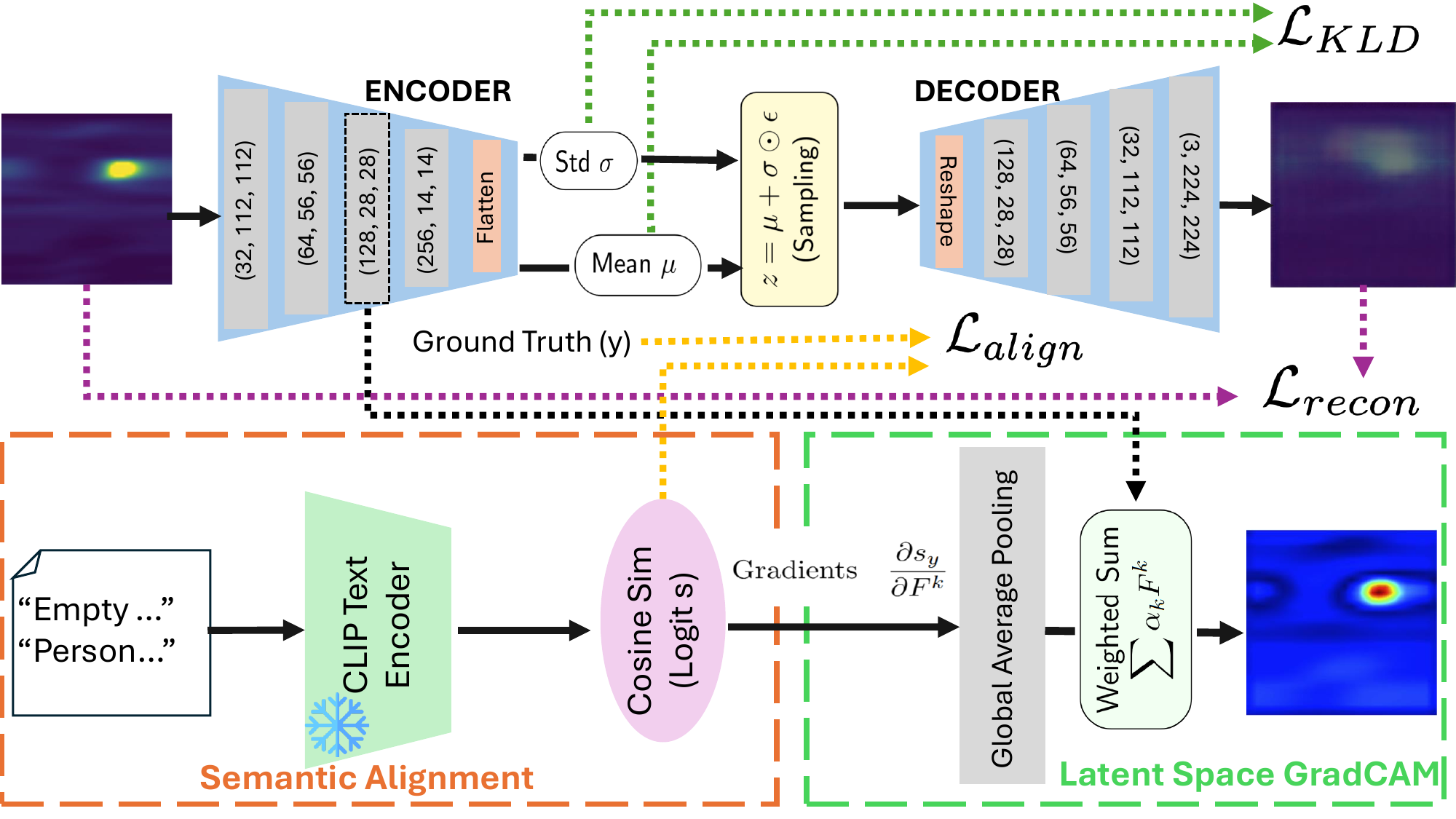} 
    \caption{The proposed Semantic Radar VAE architecture. The model learns to reconstruct the input Range-Angle heatmap while simultaneously aligning its latent representation with frozen CLIP text anchors for ``Empty'' and ``Person'' classes. This enables Latent Grad-CAM visualization of class evidence.}
    \label{fig:arch}
\end{figure}

Existing work shows that visual explanation for deep learning model predictions might appear faithful yet misaligned with human reasoning \cite{zhang2025saliencybenchcomprehensivebenchmarkevaluating}. 
Our design goal is to learn a radar representation that is simultaneously (i) generative, in the sense of capturing the underlying spatial structure of Range–Angle (RA) heatmaps, (ii) robust to noise, and (iii) semantically aligned with human concepts so that explanations can be expressed as evidence for these concepts.
To this end, we introduce a semantic radar Variational Auto Encoder (VAE) that combines a convolutional VAE backbone with a lightweight text–anchor alignment head and latent–space Grad-CAM.
An overview of the architecture is shown in Figure~\ref{fig:arch}. 
\subsection{Semantic Radar VAE}
\label{subsec:vae}
The inputs are 2D Range–Angle intensity feature maps computed from raw mmWave ADC samples using a standard FMCW processing chain: range and Doppler Fast Fourier Transform (FFT) are applied first, and angles are then estimated using Capon/Minimum Variance Distortionless Response (MVDR) beamforming \cite{10784889}.
Each processed frame is stored as an image $x \in \mathbb{R}^{C \times H \times W}$ and resized to $H \times W = 224 \times 224$, with $C=3$ channels obtained from a colormap.
Before feeding frames $x$ to the network, we scale the intensities to
$[0,1]$ and apply instance-wise normalization.
This step reduces global gain variations between frames and encourages the
VAE to focus on the geometric pattern of the RA response 
rather than absolute power \cite{10784889}.
Given input frame $x$, the encoder $E_\phi$ produces the parameters of diagonal Gaussian $q_\phi(z\mid x)$. In our implementation, it returns mean $\mu = f_\mu(x),$ and log-variance $\log\sigma^2 = f_{\log\sigma^2}(x)$ for the latent code.
The encoder consists of four strided convolutional blocks that reduce the spatial resolution from $224 
\rightarrow 14$,
followed by a flattening layer and two linear heads that produce
$\mu$ and $\log\sigma^2$ from the deepest feature maps. During training, we apply reparameterization trick \cite{Kingma_2019} to sample a latent vector $z$,
\begin{equation}
  z = \mu + \sigma \odot \epsilon,
  \quad \epsilon \sim \mathcal{N}(0, I),
  \label{eq:reparam}
\end{equation}
which allows gradients to flow through the stochastic sampling step.
This defines the approximate posterior
\begin{equation}
  q_\phi(z \mid x)
  = \mathcal{N}\bigl(z; \mu_\phi(x),
    \operatorname{diag}(\sigma_\phi^2(x))\bigr).
  \label{eq:post}
\end{equation}
The decoder $D_\theta$ consists of a stack of transposed convolutions
that upsamples $z$ back to the input resolution and outputs the reconstruction $\hat{x} = D_\theta(z)$.
We use a final sigmoid activation so that $\hat{x}$ lies in $[0,1]$
and is directly comparable to the normalized input $x$. Following the standard VAE formulation \cite{kingma2022autoencodingvariationalbayes}, the encoder and decoder are
trained to maximize the evidence lower bound (ELBO), which can be
written as the sum of a reconstruction term and a KL divergence term $L_{\text{VAE}} = L_{\text{recon}} + L_{\text{KLD}}.$ Since the RA inputs are scaled to lie in $[0,1]$, we adopt a per-pixel binary cross-entropy reconstruction loss,
\begin{equation}
  L_{\text{recon}}
  = \operatorname{BCE}(x,\hat{x})
  = -\sum_{u}
    \Bigl[x_u \log\hat{x}_u
      + (1 - x_u)\log(1-\hat{x}_u)\Bigr],
\end{equation}
where $u$ indexes pixels and channels.
And we penalize the deviation of the approximate posterior $q_\phi(z \mid x)$ from the unit Gaussian prior $p(z) = \mathcal{N}(0, I)$,
\begin{equation}
  L_{\text{KLD}}
  = D_{\mathrm{KL}}\bigl(q_\phi(z\mid x)\,\|\,p(z)\bigr)
  = -\frac{1}{2}\sum_{j}
      \bigl(1 + \log\sigma_j^2 - \mu_j^2 - \sigma_j^2\bigr),
\end{equation}
which prevents the encoder from collapsing to a set of isolated
latent codes and enables meaningful sampling and interpolation between
RA patterns.
The $L_{\text{recon}}$ encourages the model to preserve fine-grained Range–Angle structure (e.g., the shape and location of reflection angle of arrival signal return), while $L_{\text{KLD}}$ encourages a smooth, compact latent manifold that supports meaningful sampling and interpolation between RA patterns.

\subsection{Text-Anchor Latent Alignment with Frozen CLIP}
\label{subsec:clip}
To make the latent space interpretable in terms of human concepts, we
introduce a lightweight semantic alignment head that anchors the VAE
latent mean $\mu$ to text embeddings produced by a frozen CLIP text
encoder~\cite{radford2021learningtransferablevisualmodels}.  We define a set of two classes of textual prompts $T = \{t_0, t_1\}$ corresponding to these two class anchors.
Specifically, CLIP is used purely as a fixed text feature extractor, which means any other pretrained text encoder that provides a metric embedding space could be substituted without changing the rest of the method. Since the VAE latent dimension $d$ may differ from the CLIP embedding dimension $D$, we learn a linear projection $W \in \mathbb{R}^{D\times d}$
and map the latent mean as $\tilde{\mu} = W \mu.$ We then $\ell_2$-normalize the projected latent and text anchors,
\begin{equation}
  \bar{\mu} = \frac{\tilde{\mu}}{\|\tilde{\mu}\|_2}, \qquad
  \bar{t}_k = \frac{t_k}{\|t_k\|_2},
  \label{eq:norm}
\end{equation}
Each temperature-scaled cosine-similarity logits is computed as $s_k = \tau\, \bar{\mu}^\top \bar{t}_k,$ where $\tau$ is a scalar temperature parameter (initialized to $10.0$)
that controls the sharpness of the alignment.
The alignment loss is then the cross-entropy between the logits $s = [s_0, s_1]$ and $L_{\text{align}} = \operatorname{CE}(s, y)$ with $y \in \{0,1\}$ denote the ground-truth class label. This term encourages samples labelled ``empty''/``person'' to
cluster near the corresponding text anchor in the latent space, providing an intrinsic semantic structure that we later utilize for explanations. This anchor-based method is simpler than full radar-to-text generation pipelines that require additional decoders and adversarial or denoising objectives \cite{s25144467}. It is conceptually aligned with the broader motivation of projecting Gaussian latent spaces for semantic interpretability in radar while keeping the training architecture lightweight.
\subsection{Latent-Space Explainability (Latent Grad-CAM)}
To provide interpretable evidence maps, we adapt Grad-CAM~\cite{Selvaraju_2019} to the third convolutional block of the encoder. For the target class score $s_y$, we compute the gradient $\partial s_y /\partial F^c_{ij}$ with respect to the feature map $F^c \in \mathbb{R}^{H\times W}$ and derive channel important weight by global averaging pooling
\begin{equation}
    \alpha_c = \frac{1}{H \times W} \sum_{i} \sum_{j} \frac{\partial s_y}{\partial F_{ij}^c}.
\end{equation}
The latent Grad-CAM map is then
\begin{equation}
    L_{CAM} = \text{ReLU}\left(\sum_c \alpha_c F^c\right)
\end{equation}
which we upsample to the input resolution and average over Gaussian perbutations of the input to obtain a stable heatmap.

%% file: sec/experiments_and_result.tex
\begin{figure}[H]
  \centering
  \ifpdf
    \includegraphics[scale=0.32]{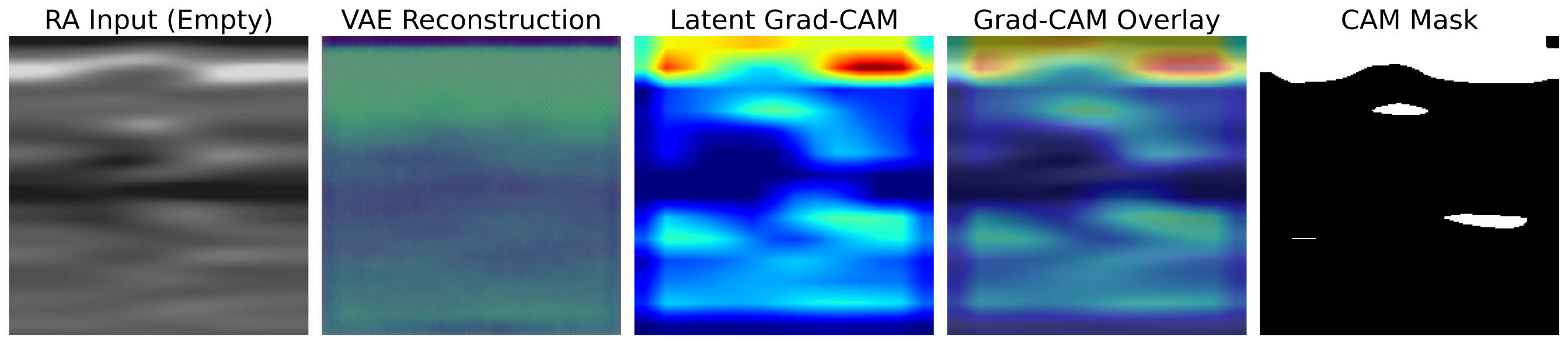}
  \else
\includegraphics[scale=0.6,natwidth=330,natheight=120]{figs/Panel_Empty_empty_room1_RIA-3Floor-UP-TV_frame_109.png.pdf}
  \fi
  \caption{Qualitative interpretability for an Empty-room RA sample using the text-aligned VAE and latent Grad-CAM.}
  \label{fig:empty_room}
\end{figure}
\begin{figure}[htb]
  \centering
  \ifpdf
    \includegraphics[scale=0.32]{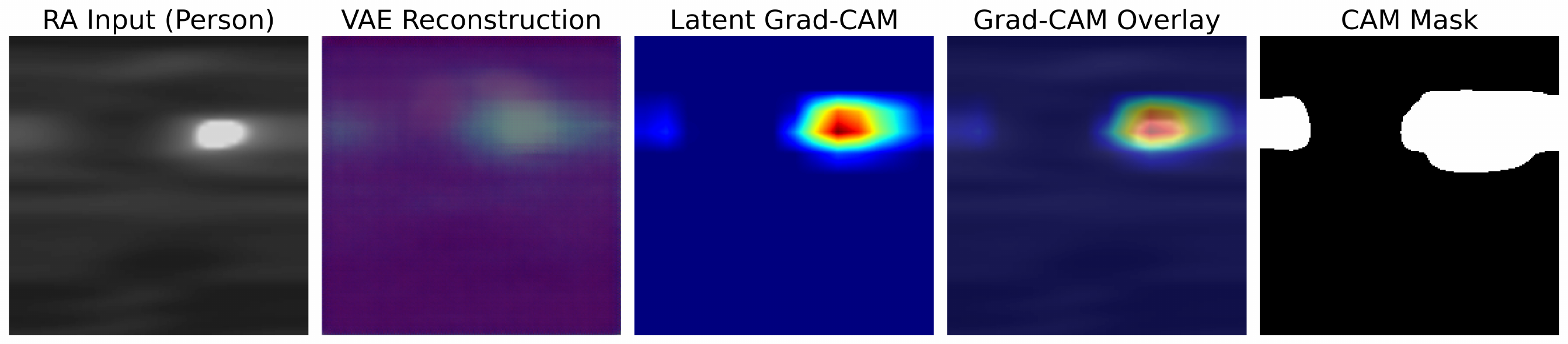}
  \else
\includegraphics[scale=0.6,natwidth=330,natheight=120]{figs/Panel_Person_Walking_Speed2_2_RIA-3Floor-UP-TV_frame_75.png.pdf}
  \fi
  \caption{Qualitative interpretability for a Person-present RA sample. The latent Grad-CAM localizes the presence-related RA blob.}
  \label{fig:person_present}
\end{figure}

We train the semantic radar VAE described in Section~\ref{sec:methodology} on the binary presence dataset of Range–Angle (RA) feature maps (``empty room'' vs.\ ``person present''). The training objective combines reconstruction, semantic alignment, and latent regularization,
$L = \lambda_r L_{\text{recon}} + \lambda_a L_{\text{align}} + \lambda_k L_{\text{KLD}},
$ with $\lambda_r = 5.0$, $\lambda_a = 1.0$, and $\lambda_k = 0.01$. These weights were chosen to keep the model generative (reconstruction dominates) while still letting the semantic anchor influence the latent geometry. 
We use Adam optimizer (learning rate $5\times10^{-4}$) and train for at most 20 epochs with early stopping (patience of 5 epochs).
The CLIP text encoder remains frozen, and gradients from $L_{\text{align}}$ update only the VAE parameters and the projection
matrix $W$.
We define two prompts corresponding to the binary presence labels
\begin{align}
  t_0 &= f_{\text{text}}\bigl(
    \text{``radar heatmap of an empty room without any person''}\bigr), \\
  t_1 &= f_{\text{text}}\bigl(
    \text{``radar heatmap of a person present in the room''}\bigr),
\end{align}
where $f_{\text{text}}$ denotes the frozen CLIP text encoder and
$t_k \in \mathbb{R}^D$ are the resulting embeddings. 
The text prompts explicitly refer to a “radar heatmap’’ and the presence or absence of a person so that the anchors are tied to the sensing modality and to the two presence classes we study. Figures~\ref{fig:empty_room} and ~\ref{fig:person_present} show qualitative results for one Empty-room and one Person-present RA sample, respectively. Each panel shows, from left to right: (i) the input RA intensity map, (ii) the VAE reconstruction, (iii) the latent Grad–CAM heatmap, (iv) the Grad–CAM overlay on the RA input, and (v) the binary CAM mask obtained by thresholding the top 15\% of CAM values. For the Empty case (Figure~\ref{fig:empty_room}), the RA input exhibits only background clutter and static multipath structure.
The VAE reconstruction is smooth and preserves the large-scale horizontal range bands, 
while the latent Grad–CAM map for the “empty room” anchor spread its energy along these bands. 
The overlay and binary mask highlight a compact high-salience region that is consistent with background reflections from furniture.
For the Person case (Figure~\ref{fig:person_present}), the RA input contains a single dominant high-intensity blob at a specific range–angle bin. The reconstruction again smooths high-frequency noise but clearly preserves the blob's position, and the latent Grad–CAM map now forms a sharply localized peak aligned with the person blob. Both the Grad–CAM overlay and the binary mask isolate this region while suppressing most of the background clutter, suggesting that the text-aligned latent space can localize presence-related structure in RA maps. In addition, the Grad-CAM explanation is qualitatively faithful to the underlying generative representation.
To probe the role of the text anchors, we perform a prompt ablation using unrelated natural-image phrases  (for e.g., ``clouds in the sky'', ``an iceberg floating in the ocean''). 
Under these random prompts, the VAE still reconstructs coarse RA patterns, but for Person-present inputs, the latent Grad-CAM becomes almost uniform and the corresponding CAM mask is empty (no clearly salient region). For Empty-room inputs, the CAM spread over the frames and no longer matches any obvious RA structure. This behaviour indicates that the semantic content of the prompts matters: radar-specific anchors are needed for the latent alignment head to produce meaningful, localized evidence maps in our setting. 

%% file: references.bib
@INPROCEEDINGS{11171388,
  author={Pal, Tapas Kumar},
  booktitle={2025 IEEE Space, Aerospace and Defence Conference (SPACE)}, 
  title={Artificial Intelligence based Radar Technology: Current Advancements, Future Directions and Challenges}, 
  year={2025},
  volume={},
  number={},
  pages={1-6},
  doi={10.1109/SPACE65882.2025.11171388}}

@ARTICLE{10554983,
  author={Kong, Hao and Huang, Cheng and Yu, Jiadi and Shen, Xuemin},
  journal={IEEE Communications Surveys \& Tutorials}, 
  title={A Survey of mmWave Radar-Based Sensing in Autonomous Vehicles, Smart Homes and Industry}, 
  year={2025},
  volume={27},
  number={1},
  pages={463-508},
  doi={10.1109/COMST.2024.3409556}}

@article{AAL,
author = {Jovanovic, Mladjan and Mitrov, Goran and Zdravevski, Eftim and Lameski, Petre and Colantonio, Sara and Kampel, Martin and Tellioglu, Hilda and Flórez-Revuelta, Francisco},
year = {2022},
month = {11},
pages = {e36553},
title = {Ambient Assisted Living: Scoping Review of Artificial Intelligence Models, Domains, Technology and Concerns},
volume = {2022},
journal = {Journal of Medical Internet Research},
doi = {10.2196/36553}
}

@article{Nahar,
author = {Nahar, Jai and Kachnowski, Stan},
year = {2023},
month = {09},
pages = {241-246},
title = {Current and Potential Applications of Ambient Artificial Intelligence},
volume = {1},
journal = {Mayo Clinic Proceedings: Digital Health},
doi = {10.1016/j.mcpdig.2023.05.003}
}

@Article{s21103549,
AUTHOR = {Cicirelli, Grazia and Marani, Roberto and Petitti, Antonio and Milella, Annalisa and D’Orazio, Tiziana},
TITLE = {Ambient Assisted Living: A Review of Technologies, Methodologies and Future Perspectives for Healthy Aging of Population},
JOURNAL = {Sensors},
VOLUME = {21},
YEAR = {2021},
NUMBER = {10},
ARTICLE-NUMBER = {3549},
URL = {https://www.mdpi.com/1424-8220/21/10/3549},
PubMedID = {34069727},
ISSN = {1424-8220},
DOI = {10.3390/s21103549}
}

@INPROCEEDINGS{10149738,
  author={Safa, Ali and Verbelen, Tim and Çatal, Ozan and Van de Maele, Toon and Hartmann, Matthias and Dhoedt, Bart and Bourdoux, André},
  booktitle={2023 IEEE Radar Conference (RadarConf23)}, 
  title={FMCW Radar Sensing for Indoor Drones Using Variational Auto-Encoders}, 
  year={2023},
  volume={},
  number={},
  pages={1-6},
  doi={10.1109/RadarConf2351548.2023.10149738}}

@article{Selvaraju_2019,
   title={Grad-CAM: Visual Explanations from Deep Networks via Gradient-Based Localization},
   volume={128},
   ISSN={1573-1405},
   url={http://dx.doi.org/10.1007/s11263-019-01228-7},
   DOI={10.1007/s11263-019-01228-7},
   number={2},
   journal={International Journal of Computer Vision},
   publisher={Springer Science and Business Media LLC},
   author={Selvaraju, Ramprasaath R. and Cogswell, Michael and Das, Abhishek and Vedantam, Ramakrishna and Parikh, Devi and Batra, Dhruv},
   year={2019},
   month=oct, pages={336–359} }

@misc{radford2021learningtransferablevisualmodels,
      title={Learning Transferable Visual Models From Natural Language Supervision}, 
      author={Alec Radford and Jong Wook Kim and Chris Hallacy and Aditya Ramesh and Gabriel Goh and Sandhini Agarwal and Girish Sastry and Amanda Askell and Pamela Mishkin and Jack Clark and Gretchen Krueger and Ilya Sutskever},
      year={2021},
      eprint={2103.00020},
      archivePrefix={arXiv},
      primaryClass={cs.CV},
      url={https://arxiv.org/abs/2103.00020}, 
}

@Article{s25144467,
AUTHOR = {Ott, Julius and Sun, Huawei and Servadei, Lorenzo and Wille, Robert},
TITLE = {How to Talk to Your Classifier: Conditional Text Generation with Radar–Visual Latent Space},
JOURNAL = {Sensors},
VOLUME = {25},
YEAR = {2025},
NUMBER = {14},
ARTICLE-NUMBER = {4467},
URL = {https://www.mdpi.com/1424-8220/25/14/4467},
PubMedID = {40732594},
ISSN = {1424-8220},
DOI = {10.3390/s25144467}
}

@INPROCEEDINGS{10784889,
  author={Abedi, Hajar and Ansariyan, Ahmad and Shaker, George},
  booktitle={2024 IEEE SENSORS}, 
  title={Contactless In-Bed Detection Using a Low-Cost Low-Resolution Radar}, 
  year={2024},
  volume={},
  number={},
  pages={1-4},
  doi={10.1109/SENSORS60989.2024.10784889}}

@misc{zhang2025saliencybenchcomprehensivebenchmarkevaluating,
      title={Saliency-Bench: A Comprehensive Benchmark for Evaluating Visual Explanations}, 
      author={Yifei Zhang and James Song and Siyi Gu and Tianxu Jiang and Bo Pan and Guangji Bai and Liang Zhao},
      year={2025},
      eprint={2310.08537},
      archivePrefix={arXiv},
      primaryClass={cs.CV},
      url={https://arxiv.org/abs/2310.08537}, 
}

@misc{lai2025radarllmempoweringlargelanguage,
      title={RadarLLM: Empowering Large Language Models to Understand Human Motion from Millimeter-Wave Point Cloud Sequence}, 
      author={Zengyuan Lai and Jiarui Yang and Songpengcheng Xia and Lizhou Lin and Lan Sun and Renwen Wang and Jianran Liu and Qi Wu and Ling Pei},
      year={2025},
      eprint={2504.09862},
      archivePrefix={arXiv},
      primaryClass={cs.LG},
      url={https://arxiv.org/abs/2504.09862}, 
}

@inproceedings{rafique2025interpretable,
  title={Interpretable Room-Level Human Presence Detection Using Ambient Sensors in Smart Homes},
  author={Rafique, Sehrish and Holthaus, Patrick and Fang, Gu and Amirabdollahian, Farshid},
  booktitle={17th International Conference on Ubiquitous Computing \& Ambient Intelligence (UCAmI 2025)},
  year={2025}
}

@ARTICLE{11071277,
  author={Alharthi, Abdullah S. and Alqurashi, Ahmed and Essa Alharbi, Turki and Alammar, Mohammed M. and Aldosari, Nasser and Bouchekara, Houssem R. E. H. and Sha’aban, Yusuf A. and Shoaib Shahriar, Mohammad and Al Ayidh, Abdulrahman},
  journal={IEEE Access}, 
  title={Explainable AI for Sensor Signal Interpretation to Revolutionize Human Health Monitoring: A Review}, 
  year={2025},
  volume={13},
  number={},
  pages={115990-116024},
  doi={10.1109/ACCESS.2025.3585764}}

@misc{kingma2022autoencodingvariationalbayes,
      title={Auto-Encoding Variational Bayes}, 
      author={Diederik P Kingma and Max Welling},
      year={2022},
      eprint={1312.6114},
      archivePrefix={arXiv},
      primaryClass={stat.ML},
      url={https://arxiv.org/abs/1312.6114}, 
}

@article{Kingma_2019,
   title={An Introduction to Variational Autoencoders},
   volume={12},
   ISSN={1935-8245},
   url={http://dx.doi.org/10.1561/2200000056},
   DOI={10.1561/2200000056},
   number={4},
   journal={Foundations and Trends® in Machine Learning},
   publisher={Emerald},
   author={Kingma, Diederik P. and Welling, Max},
   year={2019},
   pages={307–392} }
